\documentclass[twocolumn,showpacs,preprintnumbers,amsmath,amssymbm,prd]{revtex4}

\usepackage{graphicx}
\usepackage{dcolumn}
\usepackage{bm}
\usepackage{epsf}
\usepackage{amsfonts}

\newcommand{\beq}{\begin{equation}}
\newcommand{\eeq}{\end{equation}}
\newcommand{\beqr}{\begin{eqnarray} \nonumber}
\newcommand{\eeqr}{\end{eqnarray}}
\newcommand{\beqrb}{\begin{eqnarray}}
\newcommand{\eeqrb}{\nonumber \end{eqnarray}}

\newcommand{\fin}{\mbox{ .}}
\newcommand{\coma}{\mbox{ ,}}

\newcommand{\R}{{\mathbb R}}

\begin{document}

\title{Analytic Study of Rotating Black-Hole Quasinormal Modes}

\author{Uri Keshet}

\affiliation{ Institute for Advanced Study, Einstein Drive,
Princeton, NJ 08540, USA }

\altaffiliation{Friends of the Institute for Advanced Study member}

\email{keshet@sns.ias.edu}

\author{Shahar Hod}

\affiliation{The Ruppin Academic Center, Emeq Hefer 40250, Israel \\ and
The Hadassah Institute, Jerusalem 91010, Israel}

\date{\today}

\begin{abstract}
A Bohr-Sommerfeld equation is derived for the highly-damped quasinormal
mode frequencies $\omega(n\gg1)$ of rotating black holes. It may be
written as $2\int_C (p_r+ip_0) \,dr=(n+1/2)h$, where $p_r$ is the
canonical momentum conjugate to the radial coordinate $r$ along a null
geodesic of energy $\hbar \omega$ and angular momentum $\hbar m$,
$p_0=O(\omega^0)$, and the contour $C$ connects two complex turning points
of $p_r$. The solutions are $\omega(n) = - m\widehat{\omega} -
i(\widehat{\phi} + n \widehat{\delta})$, where
$\{\widehat{\omega},\widehat{\delta}\}>0$ are functions of the black-hole
parameters alone. Some physical implications are discussed.
\end{abstract}


\maketitle

Quantizing black holes may become an important step towards quantum
gravity, analogous to the role played by atomic models in the development
of quantum mechanics. Thus, the "no-hair" conjecture \cite{Wheeler71}
suggests that in a quantum theory of gravity, a black hole may be
described by few quantum numbers related to its mass $M$, electric charge
$Q$, and angular momentum $J$. The existence of classically reversible
changes in the state of a nonextremal black hole \cite{Christodoulou}
suggests that its area $A$ is an adiabatic invariant, possibly
corresponding to a quantum entity with a discrete spectrum
\cite{Bekenstein74}.

Classical black holes, like most systems with radiative boundary
conditions, are characterized by a discrete set of complex ringing
frequencies $\omega(n) = \omega_R+i\omega_I$ known as quasinormal modes
(QNMs) \cite{Nollert99}. In the spirit of Bohr's correspondence principle,
the classical QNM spectrum of a black-hole should be reproduced as
resonances in a quantum theory of gravity. QNM spectroscopy may thus
provide valuable clues towards such a theory. In particular, the
asymptotically damped frequency $\widetilde{\omega}_R \equiv
\omega_R(n\rightarrow\infty)$, which for a spherically-symmetric black
hole depends only on the black hole parameters
\cite[e.g.][]{ScRnAnalytic}, may have a simple counterpart in quantum
gravity \cite{Hod98}. Indeed, for a Schwarzschild black hole
$\widetilde{\omega}_R = (8\pi M)^{-1}\ln 3$, such that the change in black
hole entropy associated with $\Delta M=\hbar \widetilde{\omega}_R$,
$\Delta S=\Delta(4\pi M^2/\hbar)= \ln 3$, admits a (triply-) degenerate
quantum-state interpretation \cite{Hod98,SBHanalytic}. We use geometrized
units where $G=c=k_B=1$.

Although $\widetilde{\omega}$ was analytically derived for spherically
symmetric black holes \cite{SBHanalytic,ScRnAnalytic}, little is known
about the generic and more complicated case of rotating black holes.
Contradicting results for $\widetilde{\omega}$ have appeared in the
literature, although numerical convergence has recently been reported
\cite{Berti04}. An analytical solution is essential in order to test and
physically interpret these results.

We analytically derive $\widetilde{\omega}$ for rotating black holes in a
method similar to the spherical black-hole analysis of
\cite{ScRnAnalytic}, by analytically continuing the relevant solution of
Teukolsky's radial equation \cite{Teukolsky72} to the complex plane, and
matching the monodromy of the wave-function along two different contours.
Our analytical results confirm and generalize the numerical results of
\cite{Berti04}, as well as admit a physical interpretation. In this Rapid
Communication we outline the derivation and present the main results,
deferring a more elaborate description of the analysis to a future,
detailed paper.

\emph{Teukolsky's equation.--- } Linear, massless field perturbations of a
neutral, rotating black hole are described by Teukolsky's equation. For a
scalar field, this equation can be generalized to accommodate electrically
charged black holes \cite{Dudley79}; in what follows, $Q\neq 0$ is
understood to apply only to such fields. The wave-function is separated
into two ordinary differential equations using $\psi(x)=e^{i(m\phi- \omega
t)}S_{lm}(\cos \theta)R_{lm}(r)$, where $x=(t,r,\theta,\phi)$ are
Boyer-Lindquist coordinates. This yields radial and angular equations
coupled by a separation constant $A_{lm}$, where
$A_{lm}(\omega_I\rightarrow -\infty) = i A_1 a \omega+(A_0+m^2)
+O(|\omega|^{-1})$, with $A_1\in \R $ \cite{Berti04,BertiAlm}. The radial
equation then becomes
\begin{equation}
\label{eq:Teuk1} \left[ \frac{\partial^2}{\partial r^2} +
\frac{q_0(r)\omega^2 +q_1(r)\omega+ q_2(r)}{\Delta^2} \right]
\widetilde{R}_{lm} =0 \coma
\end{equation}
where $\widetilde{R}_{lm}\equiv \Delta^{(s+1)/2} R_{lm}$, $\Delta\equiv
r^2-2Mr+a^2+Q^2$, $a \equiv J/M$, and we have defined
\begin{equation}
q_0 \equiv (r^2+a^2)^2-a^2\Delta \coma
\end{equation}
\begin{eqnarray}
q_1 & \equiv & -2am(2Mr-Q^2) - i a A_1\Delta  \nonumber \\
& & +2is[r(\Delta+Q^2)-M(r^2-a^2)]  \coma
\end{eqnarray}
and
\begin{eqnarray}
q_2 & \equiv & -m^2(\Delta-a^2)-\Delta(s+A_0)  +M^2-a^2-Q^2 \nonumber \\
& & -s (M-r)[2iam+s(M-r)] \fin
\end{eqnarray}
The spin-weight parameter $s$ specifies the equation to gravitational
($s=-2$), electromagnetic ($s=-1$), scalar ($s=0$), or two-component
neutrino ($s=-1/2$) fields. For physical boundary conditions of purely
outgoing waves at both spatial infinity and the event horizon (i.e.
crossing the horizon into the black hole), Eq.~(\ref{eq:Teuk1}) admits
solutions only for a discrete set of QNM frequencies $\omega(n)$, where
$\omega_I<0$ (time decay) diverges as $n\rightarrow \infty$.

\emph{Analysis.---} By defining $z\equiv \int^r V(r')dr'$, with
$V\equiv\Delta^{-1}(q_0+\omega^{-1}q_1)^{1/2}$, Eq.~(\ref{eq:Teuk1})
becomes
\begin{equation}
\label{eq:Teuk2} \left(-\frac{\partial^2}{\partial
z^2}+V_1-\omega^2\right)\widehat{R} = 0 \coma
\end{equation}
where $\widehat{R}=V^{1/2} \widetilde{R}$ and $V_1 =
V''/(2V^3)-3(V')^2/(4V^4) - q_2/(V\Delta)^2$. A nonconventional tortoise
coordinate $z$ was defined such that the effective potential
$V_1=O(|\omega|^0)$. The boundary condition at the horizon becomes
$\widehat{R}(r\rightarrow r_+)\sim \exp{(-i\omega z)}\propto
(r-r_+)^{-i\omega\sigma_+}$, where
\begin{eqnarray}
\label{eq:SigmaPlus} \omega\sigma_+ =  \omega \underset{\,\,\,r\rightarrow
r_+}{\mbox{Res}} (V) = \beta \left(\omega -m \Omega\right) - \frac{is}{2}
+ O(|\omega|^{-1}) \fin
\end{eqnarray}
Here, $\Omega\equiv a/(r_+^2+a^2)$ is the angular velocity of the event
horizon, $\beta\equiv \hbar/(4\pi T)=(r_+^2+a^2)/(r_+-r_-)$, $T$ is the
Bekenstein-Hawking temperature, $r_\pm=M\pm (M^2-a^2-Q^2)^{1/2}$ are the
outer and inner horizon radii, and the tilde in $\widetilde{\omega}$ is
omitted unless necessary (henceforth). $\widehat{R}(r\simeq r_+)$ is
multivalued, such that a clockwise rotation around $r_+$ multiplies
$\widehat{R}$ by a factor $\Phi_1 = \exp{(-2\pi \omega\sigma_+)}$.

\begin{figure}[h]
\centerline{\epsfxsize=7.5cm \epsfbox{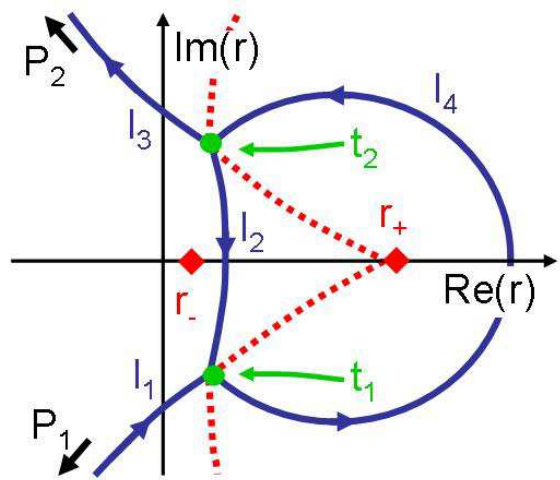}}
\caption{\label{fig:Stokes} Illustration of anti-Stokes (solid) and Stokes
(dashed) lines emanating from the turning points $t_1$ and $t_2$ (disks)
in the complex $r$-plane, for $a=0.3$, $Q=0$ in the highly damped limit.
The inner and outer horizon radii (diamonds) and components of the contour
$C$ are also shown. Arrows along anti-Stokes lines denote the direction of
increasing $\Im z$.}
\end{figure}

Let $r_1$ and $r_2=r_1^*$ be the two complex conjugate roots of $q_0(r)$
lying in the fourth and in the first quadrants, respectively. Denote $t_1$
and $t_2$ as the turning points of $V$ [defined by $V(r=t_i)=0$] which lie
near (a factor $\sim |\omega|^{-1}$ away from) $r_1$ and $r_2$,
respectively (see Figure \ref{fig:Stokes}). The monodromy $\Phi_2$ of
$\widehat{R}$ along a clockwise contour $C$, which passes through $t_1$
and $t_2$ and encloses $r_+$, is used to determine $\omega$ by demanding
$\Phi_1=\Phi_2$, as in \cite{ScRnAnalytic}. A reader uninterested in
details of the derivation may skip directly to the result,
Eq.~(\ref{eq:QNM1}).

Near the turning points, $(z-z_i)\propto (r-t_i)^{3/2}$, where $z_i\equiv
z(t_i)$. Therefore three anti-Stokes lines, defined by $\Re (i\omega
z)=0$, emanate from $t_i$. Two anti-Stokes lines connect $t_1$ to $t_2$;
one (denoted $l_2$) crosses the real axis between $r_-$ and $r_+$, while
the other crosses it at $r>r_+$. The third anti-Stokes line ($l_1$)
emanating from $t_1$ extends to $P_1$, where $|P_1|\rightarrow \infty$ and
$\arg (P_1)=-\pi/2$. A similar line ($l_3$) runs from $t_2$ to $P_2$, with
$|P_2|\rightarrow \infty$ and $\arg (P_2)=+\pi/2$. A Stokes line, defined
by $\Im (i\omega z)=0$, emanates between every two anti-Stokes lines of
$t_i$. Let $C$ be the closed, clockwise contour running from $P_1$ to
$P_2$ along the anti-Stokes lines $l_1$, $l_2$ and $l_3$, and closing back
on $P_1$ through the large semicircle $l_\infty$, where
$|r|\rightarrow\infty$ and $-\pi/2 < \arg(r) < \pi/2$. The turning points
$t_1$ and $t_2$ are excluded from $C$ by partially rotating around them
counterclockwise. Figure \ref{fig:Stokes} illustrates these features in
the $r$-plane.

Along anti-Stokes lines, the WKB approximation $\widehat{R}(z,z_0)\simeq
c_+ \exp{[+i \omega (z-z_0)]} +c_- \exp{[-i \omega (z-z_0)]}$ holds. Off
the lines, this may also be written as $c_d f_d+c_s f_s$, where $f_d$ is
exponentially large (dominant) and $f_s$ is exponentially small
(subdominant). For $\omega_R<0$, the boundary condition at spatial
infinity can be analytically continued to $P_1$ \cite{ScRnAnalytic} such
that $\widehat{R}(P_1)\sim \exp{(+i\omega z)}$, i.e. $\{c_+,c_-;z_0\} =
\{1,0;z_1\}$ up to a multiplicative factor. This remains invariant along
$l_1$ till the vicinity of $t_1$, so we denote
$\widehat{R}(l_1)=\{1,0;z_1\}$. When an anti-Stokes line is crossed, the
dominant and subdominant parts exchange roles. When a Stokes line is
crossed while circling a regular turning point, $c_d f_d+c_s f_s$ becomes
$c_d f_d+(c_s\pm i c_d)f_d$, where the positive (negative) sign
corresponds to a counterclockwise (clockwise) rotation. This so-called
Stokes phenomenon \cite{StokesPhenomenon} implies that after rotating
around $t_1$ from $l_1$ to $l_2$, thus crossing two Stokes lines and the
anti-Stokes line between them, $\widehat{R}(l_2)=\{0,i;z_1\} = \{0,i\exp
(-i\omega\delta);z_2\}$, where
\begin{equation} \label{eq:deltaDef}
\delta \equiv z_2-z_1 = \int_{l_2} V \, dr \fin
\end{equation}
Similarly, after rotating from $l_2$ to $l_3$, $\widehat{R}(l_3) =
\{-\exp(-2i\omega\delta),0;z_1\}$. Finally, along $l_\infty$ the
coefficient of the dominant part of the solution $c_+$ remains invariant
till $P_1$. In addition to the above changes in $c_+$, it accumulates a
phase $e^{+2\pi\omega\sigma_+}$ due to the (only) singularity at $r_+$
enclosed by $C$. Thus, the total phase accumulated by $\widehat{R}$ along
$C$ is $\Phi_2=-\exp(-2i\omega\delta+2\pi\omega\sigma_+)$. For
$\omega_R>0$, the boundary condition at spatial infinity is continued to
$P_2$ and the two contours are chosen counterclockwise, such that the
resulting equation $\Phi_1=\Phi_2$ is unchanged.

The constraint $\Phi_1=\Phi_2$ finally yields the highly-damped QNM
equation \footnote{Eq.~(\ref{eq:QNM1}) can also be derived as in Ref.
\cite{ScRnAnalytic}, by solving for $\widehat{R}$ near the the turning
points where $V_1 \simeq -(5/36)(z-z_i)^{-2}$.}
\begin{equation}
\label{eq:QNM1}
e^{-2\pi\omega\sigma_+}=-e^{-2i\omega\delta+2\pi\omega\sigma_+}\fin
\end{equation}
Explicitly, to order $O(|\omega|^{-1})$ this may be written as
\begin{equation}
\label{eq:QNM2} 4\pi \beta \left(\omega -m \Omega\right) - 2\pi i s =
2i\omega \int_{C_{t,i}} V \, dr -\pi i(2n+1) \coma
\end{equation}
or in a more compact form as
\begin{equation}
\label{eq:QNM2B} 2 \omega \int_{C_{t,o}} V \, dr = 2\pi \left(
n+\frac{1}{2} \right) \coma
\end{equation}
where $n\in{\mathbb Z}$. Here, $C_{t,i}$ ($C_{t,o}$) is a complex-plane
contour running from $t_1$ to $t_2$, crossing the real axis in (out) of
the event horizon, at some point $r_-<r<r_+$ ($r>r_+$).

Before solving for $\widetilde{\omega}$, note that in the highly-damped
limit the real and the imaginary contributions to the integrals of
Eqs.~(\ref{eq:deltaDef})-(\ref{eq:QNM2B}) are easily separated. For
example, the real part of Eq.~(\ref{eq:QNM2}) may be written in the form
\footnote{Using $\int_{r_1}^{r_2} i|f|dr\in\R$. The integration endpoints
$\{t_i\}$ and $\{r_i\}$ may be used interchangeably, as $q_0(t_i)=0$
ensures that the resulting $O(|\omega|^{-1})$ correction terms vanish.}
\begin{equation} \label{eq:omegaReff}
4\pi\beta (\omega_R-m \Omega) = \Re{ \left(2i\int_{C_{t,i}} \omega V_R\,dr
\right)} \coma
\end{equation}
where the complex potential $V_R$ is given by
\begin{equation} \label{eq:omegaRVeff}
(\omega V_R)^2 = \frac{ q_0\omega^2 -2am(2Mr-Q^2)\omega -
m^2(\Delta-a^2)}{\Delta^2} \fin
\end{equation}
The last term ($\propto \omega^0$, taken from $q_2$) was added to $V_R$
for future use and has no effect in the highly-damped limit. An equation
analogous to Eq.~(\ref{eq:omegaReff}) is found for the imaginary part
$4\pi\beta \omega_I - 2\pi s$.

\emph{QNM frequencies.---} In order to obtain a closed-form expression for
$\omega$, expand $2i\delta - 4\pi \sigma_+= \delta_0+ (m\delta_m+
is\delta_s+ iA_1 \delta_A)\omega^{-1} + O(|\omega|^{-2})$. Here
\begin{equation}
\delta_j \equiv 2i\int_{C_{r,o}} V_j \,dr \coma
\end{equation}
with $V_0 = q_0^{1/2}\Delta^{-1}$, $V_m = -a
(2Mr-Q^2)\Delta^{-1}q_0^{-1/2}$, $V_s =
[r(\Delta+Q^2)-M(r^2-a^2)]\Delta^{-1}q_0^{-1/2}$, and $V_A =
-q_0^{-1/2}a/2$. The integration contour $C_{r,o}$ runs from $r_1$ to
$r_2$, crossing the real axis outside the event horizon. Since
$r_2=r_1^*$, $\{\delta_0,\delta_s, \delta_A, \delta_m\}$ are all real.
Analytic expressions for these $\delta_j$ functions are readily found in
terms of elliptic integrals.

\begin{figure}[h]
\centerline{\epsfxsize=7.9cm \epsfbox{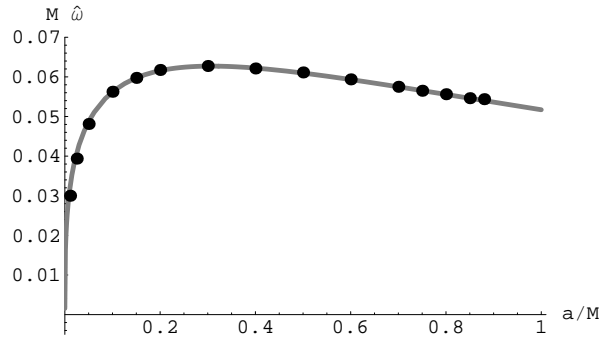}}
\caption{\label{fig:omegaR} The real part of the highly damped QNM
frequency $\widehat{\omega}(a) = \widetilde{\omega}_R(a;m=-1)$ for $Q=0$,
according to Eq.~(\ref{eq:QNM3}) (line) and according to the numerical
results of \cite{Berti04} (circles).}
\end{figure}

\begin{figure}[b]
\centerline{\epsfxsize=7.9cm \epsfbox{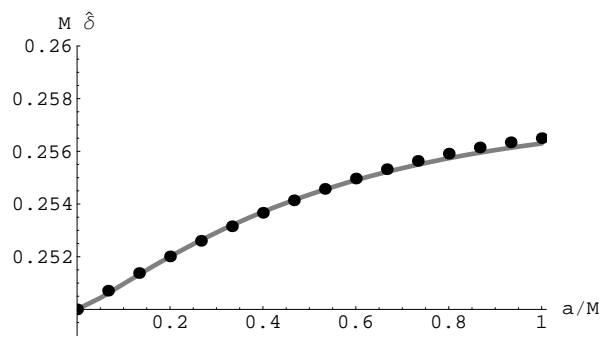}}
\caption{\label{fig:omegaI} Level spacing
$|\Delta\omega(a)|=\widehat{\delta}$ for $Q=0$ according to
Eq.~(\ref{eq:QNM3}) (line) and the numerical fit in \cite{Berti04}
(circles).}
\end{figure}

With the above definitions we finally obtain
\begin{equation} \label{eq:QNM3}
\omega=-m\widehat{\omega}-i(\widehat{\phi}+n\widehat{\delta})\coma
\end{equation}
where $\widehat{\omega}=\delta_m/\delta_0$, $\widehat{\delta} = 2\pi
/\delta_0$, and $\widehat{\phi} = (s\delta_s + A_1\delta_A-\pi)/\delta_0$.
As shown in Figures \ref{fig:omegaR} and \ref{fig:omegaI}, these analytic
results agree with the numerical calculations of \cite{Berti04}.

Eq.~(\ref{eq:QNM3}) yields one branch of solutions $\omega_m(n)$ in the
asymptotic limit. Interestingly, in the low-$n$ regime (and in
spherically-symmetric black holes) two branches of solutions are
identified, for given field and black-hole parameters \cite{Leaver85}.

The asymptotic QNMs are not continuous at $a=0$ \footnote{The analysis is
valid only for $0<a^2<M^2-Q^2$. It does not apply for $a=0$, where $r_1$
and $r_2$ coalesce to $0$, nor in the extremal case $M^2-a^2-Q^2=0$, where
$r_-$ and $r_+$ merge to cut off the anti-Stokes line $l_2$. It does apply
in the extremal limit, where numerical calculations fail and we find
$\widehat{\omega}(a\rightarrow M) \simeq 0.051704/M$.}. For $Q=0$,
$\widehat{\omega}(a\rightarrow 0) \propto a^{1/3}\rightarrow 0$, whereas
$\omega_R(a=0) = (8\pi M)^{-1}\ln3$. Such discontinuous behavior sometimes
occurs in the Schwarzschild limit, for example in the inner structure of
the black hole \cite{InnerStructure}. Note that the level spacing
$\widehat{\delta}$ does continuously asymptote to the Schwarzschild result
$\Delta\omega = 2\pi T/\hbar$ \cite{SBHanalytic} as $\{a,Q\}\rightarrow
0$.

\emph{Discussion.---} We have analytically studied the highly-damped QNM
frequencies $\omega(n)$ of a rotating black hole. A Bohr-Sommerfeld-like
equation for $\omega$ was derived [Eqs.~(\ref{eq:QNM2})-(\ref{eq:QNM2B})],
analytically solved [Eq.~(\ref{eq:QNM3})], and shown to agree and
generalize previous numerical results \cite{Berti04} (Figures
\ref{fig:omegaR} and \ref{fig:omegaI}).

It is instructive to quantize the linear field perturbations described by
the QNM \footnote{The analysis can alternatively proceed in the
geometrical optics approximation, where radiation follows null
geodesics.}. A quantum of complex energy $\hbar\omega(n)$ and angular
momentum $\hbar m$ may thus be associated with the highly-damped QNM
frequency $\omega_m(n)$. Multiplying Eq.~(\ref{eq:QNM2B}) by $\hbar$
yields
\begin{equation}
\label{eq:BohrSommerfeld1} 2 \int_{C_{t,o}} p \, dr = \left( n+\frac{1}{2}
\right)h \coma
\end{equation}
where $p=\hbar \omega V$. This equation strongly resembles the
Bohr-Sommerfeld quantization rule $\oint p\,dq=(n+1/2)h$, where $p$ is the
canonical momentum conjugate to some coordinate $q$, and the integration
is carried out along a closed orbit. To elucidate the connection, recall
that the covariant radial momentum $p_r$ for geodesic motion of a neutral,
massless particle of energy $E$ and angular momentum $p_\phi$, is given by
\begin{eqnarray}
(p_r \Delta)^2 & = & [(r^2+a^2)^2-a^2\Delta]E^2 -
2a(2Mr-Q^2)Ep_\phi \nonumber \\
& &   - (\Delta-a^2)p_\phi^2-Q_C\Delta \coma
\end{eqnarray}
where $Q_C$ is Carter's (fourth) constant of motion \cite{Carter68}.
Comparing this with Eq.~(\ref{eq:omegaRVeff}) indicates that $V_R \approx
p_r$, provided that $E=\hbar \omega$, $p_\phi=\hbar m$, and $Q_C=O(E^0)$.
Hence, up to an $O(\omega^0)$ term which leads to an imaginary offset in
$\omega(n)$, the integrand in Eq.~(\ref{eq:BohrSommerfeld1}) truly is of
the form $p\,dq$ for the above QNM quantization. The implied physical
content of Eq.~(\ref{eq:BohrSommerfeld1}) suggests that the full QNM
spectrum may be determined by a generalized Bohr-Sommerfeld equation,
which reduces to Eq.~(\ref{eq:BohrSommerfeld1}) as
$\omega_I\rightarrow-\infty$. The general form of $p$ is not uniquely
determined by our highly-damped analysis. Up to $O(|\omega|^{-1})$
corrections, we may write
\begin{equation} \label{eq:pASpr}
p=p_r+i\hbar s V_s+i\hbar A_1 V_A \, .
\end{equation}

The preceding discussion implies that Eq.~(\ref{eq:BohrSommerfeld1}) can
be interpreted as a complex version of the Bohr-Sommerfeld quantization
rule. This rule was used in (the old) quantum mechanics to determine the
quantum-mechanically allowed trajectories, as well as the quantized values
of the associated constants of motion. Realizing the full meaning of
Eq.~(\ref{eq:BohrSommerfeld1}) may well require a quantum theory of
gravity. Conversely, this equation can possibly be used to constrain and
shed light on the theory.

The quantum manifestation of a QNM may be complicated. A simple example is
motivated by the outgoing boundary conditions of the QNMs and the symmetry
of their frequencies $\omega_{-m} = -\omega_m^*$ \cite{Leaver85}, evident
in Eq.~(\ref{eq:QNM3}). These suggest that a quantum pair of opposite
angular momentum may fundamentally correspond to a QNM; a positive energy
quantum escaping to infinity and a negative energy quantum falling into
the black hole, in resemblance of Hawking's semiclassical radiation. Under
such circumstances, a quantum process corresponding to a QNM changes the
black-hole mass by $\Delta M=\hbar \omega_R$ and its angular momentum by
$\Delta J=\hbar m$. For such small changes in the black-hole parameters,
the corresponding change in its entropy, $\Delta S=T^{-1}(\Delta M -
\Omega\Delta J)$, is given directly by Eq.~(\ref{eq:omegaReff}), which we
may now write as
\begin{equation} \label{eq:DeltaA}
\hbar \Delta S = \Delta A/4 = \Re\left(2i\int_{C_{t,i}} p_r \,dr\right)
\fin
\end{equation}
This is another indication of the adiabatic invariance of the area/entropy
\cite{Bekenstein74}.

We thank A. Neitzke, J. Maldacena, P. Goldreich and J. Bekenstein for
helpful discussions. U.K. is supported by the NSF (grant PHY-0503584).

\end{document}